# Measurement of In-Circuit Common-Mode Impedance at the AC Input of a Motor Drive System

Zhenyu Zhao, Fei Fan, Arjuna Weerasinghe, Pengfei Tu, and Kye Yak See
School of Electrical and Electronic Engineering
Nanyang Technological Univeristy
Singapore
zhao0245@e.ntu.edu.sg

*Abstract*—The in-circuit common-mode (CM) impedance at the AC input of a motor drive system (MDS) provides valuable inputs for evaluating and estimating the CM electromagnetic interference (EMI) noise generated by the switching of power semiconductor devices in the MDS. This paper introduces a single-probe setup (SPS) with frequency-domain measurement to extract the in-circuit CM impedance of a MDS under its different operating modes. The SPS has the merits of non-contact measurement and simple structure.

## I. INTRODUCTION

The switching of power semiconductor devices in a motor drive system (MDS) will generate conducted electromagnetic interference (EMI) noise, which can propagate into the power grid from its AC input, and consequently affects the normal operation of other grid-connected electrical assets [1]. For conducted EMI noise, it is usually separated into common-mode (CM) and differential-mode (DM) components. To evaluate and estimate these noise components, the respective CM and DM equivalent noise models of the MDS can be constructed [2], [3]. Since these noise models are usually represented by the respective CM and DM equivalent noise sources with internal impedances, it is necessary to extract these internal impedances for noise evaluation and estimation. Compared with off-circuit impedance measurement, in-circuit impedance measurement will bring more accurate and realistic results because the impedances of the MDS under its actual operating conditions can be very different from those when the MDS is off-circuit [4].

Many in-circuit impedance measurement methods have been reported, which can be generalized into three categories: the voltage-current (V-I) measurement approach [5]-[8], the capacitive coupling approach [9]-[11], and the inductive coupling approach [12]-[14]. The V-I measurement approach extracts the in-circuit impedance of an energized electrical system under test (SUT) by using voltage and current sensors, in which the voltage senor is used to measure the test signal voltage cross the SUT and the current sensor is used to measure the test signal current flowing through the SUT [5]. The test signal of this approach can be either the existing harmonics present in the SUT [6] or an externally injected signal [7]. Based on the obtained test signal voltage and current, combined with the digital signal processing (DSP) algorithm, the in-circuit impedance of the energized SUT can be determined. The capacitive coupling approach extracts the in-circuit impedance of an energized SUT by using a few coupling capacitors with a vector network analyzer (VNA) [9] or an impedance analyzer (IA) [10]. The coupling capacitors are connected between the VNA/IA and the energized SUT to isolate the DC or low-frequency AC power source but provide a low-impedance path for the high-frequency test signal that is generated by the VNA/IA [11]. The inductive coupling approach extracts the in-circuit impedance of an energized SUT usually by using two clamp-on inductive probes with a measurement instrument, in which one of the inductive probes is used for injecting an excitation test signal (generated by the measurement instrument) into the SUT and the other inductive probe is used for receiving the response of the same test signal. By creating the relationship between the excitation and response test signals, the in-circuit impedance of the energized SUT can be confirmed [12]. All the three approaches have been well established for specific applications. However, the voltage sensor used in the V-I measurement approach and the coupling capacitors used in the capacitive coupling approach require direct electrical contact with the energized SUT for in-circuit measurement, which demands special provision for such contact, especially when the SUT is energized by high voltages. In contrast, the measurement setup of the inductive coupling approach has no direct electrical contact with the energized SUT and therefore it simplifies the on-site implementation without posing electrical safety hazards.

The inductive coupling approach was first proposed for power line in-circuit impedance measurement [12]. Then, it was improved and extended to many other applications [3], [4], [13]. The classic measurement setup (called "two-probe setup") of this approach is realized by two clamp-on inductive probes and a frequency-domain measurement instrument, such as a VNA with sweep frequency excitation [13]. Also, several recent developments on the two-probe setup (TPS) have been carried out by using a time-domain measurement instrument for realizing time-variant in-circuit impedance monitoring [14]. However, the TPS with time-domain measurement can only monitor the in-circuit impedance at one frequency at time, which is time-consuming for measuring the in-circuit impedance response in a wide frequency range in comparison to the TPS with frequency-domain measurement. Regardless of with frequency-domain measurement or time-domain measurement, the TPS suffers from the probe-to-probe coupling between the two inductive probes, which can deteriorate its measurement accuracy. Although a calibration technique for the TPS with time-domain measurement has been proposed to de-embed the effects of probe-to-probe coupling on the measurement accuracy, it still cannot eliminate this coupling fundamentally [15].

To eliminate the probe-to-probe coupling, a single-probe setup (SPS) with frequency-domain measurement has been recently developed [16]. The SPS basically consists of a clamp-on inductive probe and a VNA. It can also be incorporated a signal amplification and protection (SAP) module to improve its signal-to-noise ratio (SNR) and enhance its ruggedness to make it a good candidate even for high power applications with significant background noise (e.g. harmonics) and power surges. Considering the aforesaid advantages, this paper discusses the practical application of the SPS to measure the in-circuit CM impedance at the AC input of a MDS under the MDS's different operating modes.

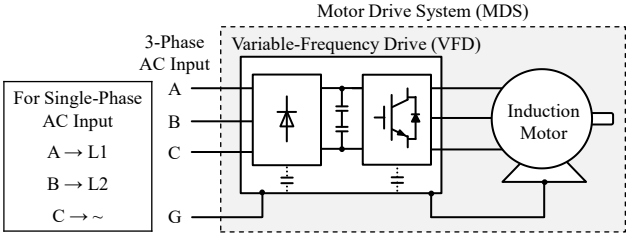

Fig. 1. Schematic diagram of a typical MDS.

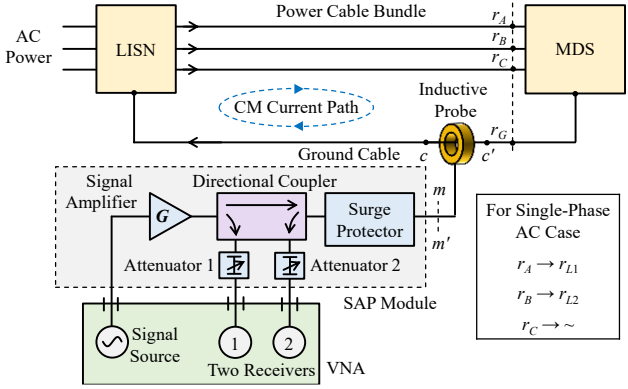

Fig. 2. Experimental setup of measuring the in-circuit CM impedance at the AC input of the MDS through the SPS.

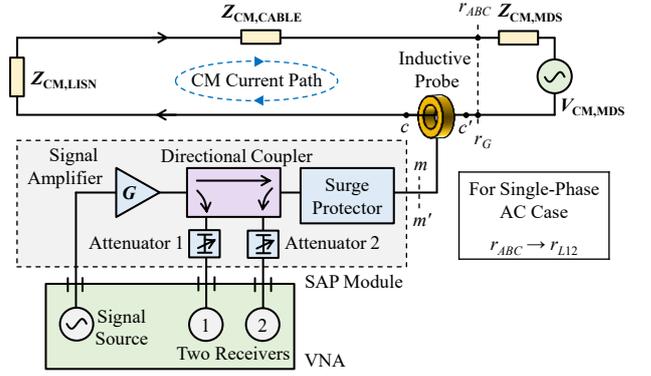

Fig. 3. CM equivalent circuit of Fig. 2.

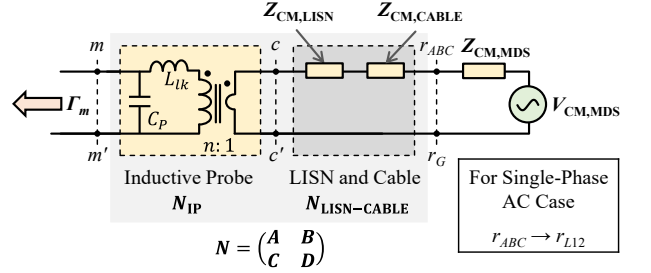

Fig. 4. Cascaded two-port networks representation of Fig. 3.

This paper is organized as follows. Section II details the use of the SPS for measuring the in-circuit CM impedance at the AC input of a MDS. Using a commercial available MDS as a test case, Section III shows the measured in-circuit impedances of the MDS under its different operating modes. Finally, Section IV concludes this paper.

## II. MEASUREMENT OF IN-CIRCUIT CM IMPEDANCE AT THE AC INPUT OF A MDS

Fig. 1 shows a schematic diagram of a typical MDS, which consists of a variable frequency drive (VFD) and an induction motor with cables in between. The AC input of the MDS is usually 3-phase or single-phase. Fig. 2 shows the block diagram of measuring the in-circuit CM impedance at the AC input of the MDS ($Z_{CM,MDS}$) through the SPS, in which the MDS is connected to the AC power through a line impedance stabilization network (LISN). The presence of the LISN is to provide a stable and well-defined impedance at the AC power side [4]. The switching of power semiconductor devices in the VFD generates CM noise at the AC input of the MDS, which results in a CM current path formed by the MDS, power cable bundle, LISN, and ground cable. As shown in Fig. 2, the SPS consists of a VNA, a clamp-on inductive probe, and a SAP module. The SAP module is applied because the MDS usually generates significant background noise (e.g. harmonics) and experiences power surges. For in-circuit measurement of $Z_{CM,MDS}$, the inductive probe is clamped on the ground cable with the position marked as $c$-$c'$.

Fig. 3 shows the CM equivalent circuit of Fig. 2, where $V_{CM,MDS}$ denotes the equivalent CM noise voltage source of the MDS; $Z_{CM,CABLE}$ denotes the equivalent CM loop impedance formed by the power cable bundle and the ground cable; $Z_{CM,LISN}$ denotes the equivalent CM impedance of the LISN. For measuring in-circuit $Z_{CM,MDS}$, a sweep-frequency test signal is generated by the signal source of the VNA. The test signal is amplified by the signal amplifier and then injected into the CM path through the inductive probe. By using the directional coupler, the incident and reflected waves of the test signal can be separated and respectively measured by the two receivers of the VNA. The two attenuators are added to ensure the power of the measured test signal within the allowable input range of the receivers. The surge protector is used to protect the VNA from power surges in the MDS.

Based on the network analysis theory, Fig. 4 shows the cascaded two-port networks representation of Fig. 3 from $m$-$m'$ [16]. $\Gamma_m$ is the reflection coefficient observed at $m$-$m'$, which is calculated directly by the VNA using the incident and reflected waves of the test signal measured by the two receivers [17]. $N_{IP}$ is the two-port network of the inductive probe, in which $L_{lk}$ and $C_p$ denote its leakage inductance and equivalent parasitic capacitance, respectively. $N_{LISN\text{-}CABLE}$ is the two-port network of the LISN and cables. Since $N_{IP}$ and $N_{LISN\text{-}CABLE}$ are cascaded, the resulting two-port network $N$ can be expressed as:

$$N = N_{IP} \cdot N_{LISN-CABLE} \quad (1)$$

From Fig. 4, the relationship between $Z_{CM,MDS}$ and $\Gamma_m$ can be established in terms of the transmission (*ABCD*) parameters of $N$ as follows:

$$Z_{CM,MDS} = \frac{k_1 \cdot \Gamma_m + k_2}{\Gamma_m + k_3} \quad (2)$$

where

$$k_1 = -\frac{Z_0 \cdot D + B}{Z_0 \cdot C + A} \quad (3)$$

$$k_2 = -\frac{Z_0 \cdot D - B}{Z_0 \cdot C + A} \quad (4)$$

$$k_3 = \frac{Z_0 \cdot C - A}{Z_0 \cdot C + A} \quad (5)$$

where $Z_0$ is the reference impedance of the VNA.

From (2), once $k_1$, $k_2$, and $k_3$ are known, $Z_{CM,MDS}$ can be determined by the extracted $\Gamma_m$ from the VNA. By observing (3)-(5), the frequency-dependent $k_1$, $k_2$, and $k_3$ are determined by $ABCD$ parameters of $N$ and $Z_0$, which remain unchanged for given SPS, LISN, and cables. To determine $k_1$, $k_2$, and $k_3$, a pre-measurement characterization procedure has been well reported in [16] and will only be briefly described here. For performing the characterization, the open, short, and 50-$\Omega$ load conditions are respectively realized at the position $r_{ABC}$-$r_G$. Since the CM impedances of the LISN and the cables maintain unchanged regardless of the "on" or "off" condition of AC power, the characterization is performed when the AC power is "off" condition. Based on the characterization, $k_1$, $k_2$, and $k_3$ can be finally determined as follows:

$$k_1 = 50 \cdot \frac{\Gamma_L - \Gamma_O}{\Gamma_L - \Gamma_S} \quad (6)$$

$$k_2 = 50 \cdot \Gamma_S \cdot \frac{\Gamma_O - \Gamma_L}{\Gamma_L - \Gamma_S} \quad (7)$$

$$k_3 = -\Gamma_O \quad (8)$$

where $\Gamma_O$, $\Gamma_S$, and $\Gamma_L$ represent the measured reflection coefficients by the VNA respectively at the open, short, and 50-$\Omega$ load conditions at $r_{ABC}$-$r_G$.

## III. EXPERIMENT

In this section, a commercially available MDS is used as a test case. The specifications of the MDS, LISN and cables are listed in Table I, and Table II gives the details of the SPS used for measuring the in-circuit $Z_{CM,MDS}$. The pre-measurement characterization procedure is carried out and Fig. 5 shows the determined $k_1$, $k_2$, and $k_3$ of the selected setup from 150 kHz to 30 MHz. After obtaining $k_1$, $k_2$, and $k_3$, in-circuit $Z_{CM,MDS}$ measurement can be carried out through the SPS.

$Z_{CM,MDS}$ is measured under six operating modes of the MDS to study the characteristics of in-circuit $Z_{CM,MDS}$ in each mode. Since the selected VFD supports voltage/frequency (V/F) and sensorless-vector (SLV) control modes, and the rated frequency of the induction motor is 50 Hz, the details of the operating modes are listed in Table III. Fig. 6(a) shows the measurement results under the six operating modes of the MDS. For clarity, Figs. 6(b)-(f) show a series of comparisons of the measurement results among the operating modes. From Figs. 6(b)-(d), at the same VFD's output frequency, the in-circuit $Z_{CM,MDS}$ values under V/F control and SLV control show very good consistency over the entire frequency range of 150 kHz to 30 MHz. It indicates that the control modes have negligible influence on the in-circuit $Z_{CM,MDS}$ value. In contrast, Figs. 6(e)-(f) show that the VFD's output frequency setting affects the in-circuit $Z_{CM,MDS}$ value in certain frequency regions, although this influence is rather small.

TABLE I. SPECIFICATIONS OF THE MDS, LISN, AND CABLES

| Component | Specifications |
| --- | --- |
| VFD | TECO L510s (No built-in EMI filter) |
| Induction Motor | RMS8024/B3 (3 phase, 4 pole, 0.75 kW, 50 Hz) |
| LISN | Electro-Metrics MIL 5-25/2 (100 kHz-65 MHz) |
| Cables | VFD to Induction Motor: 60 cm<br>MDS to LISN: 100 cm |

TABLE II. DETAILS OF THE SPS

| Component | Details |
| --- | --- |
| Inductive Probe | SOLAR 9144-1N (4 kHz-100 MHz) |
| VNA | Omicron Bode 100 |
| Signal Amplifier | Mini Circuits LZY–22+ (100 kHz-200 MHz) |
| Directional Coupler | DC3010A (10 kHz-1 GHz) |
| Surge Protector | SSC-N230/01 |
| Attenuator 1 | AIM-Cambridge 27-9300-6 (6 dB) |
| Attenuator 2 | AIM-Cambridge 27-9300-3 (3 dB) |

TABLE III. OPERATING MODES OF THE MDS

| Operating Mode | Control Mode | VFD's Output Frequency |
| --- | --- | --- |
| Mode 1 | V/F Control | 10 Hz |
| Mode 2 | V/F Control | 30 Hz |
| Mode 3 | V/F Control | 50 Hz |
| Mode 4 | SLV Control | 10 Hz |
| Mode 5 | SLV Control | 30 Hz |
| Mode 6 | SLV Control | 50 Hz |

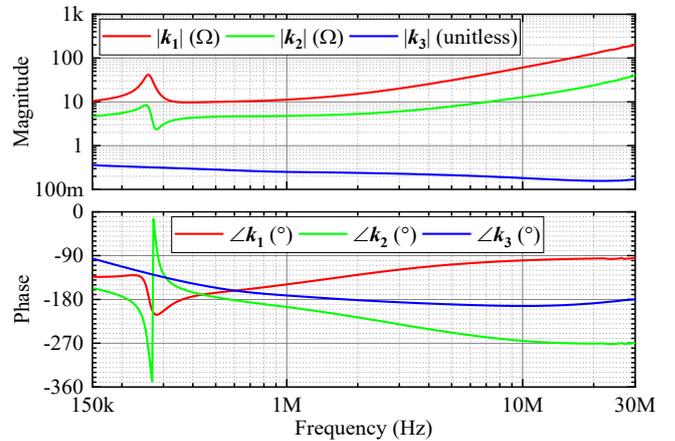

Fig. 5. Determined $k_1$, $k_2$, and $k_3$ from 150 kHz to 30 MHz.

## IV. CONCLUSION

This paper discusses the practical application of the SPS to extract the in-circuit CM impedance of a MDS under its different operating modes. By using a commercially available MDS as a test case, it has been demonstrated experimentally that the control modes of the MDS have negligible impact on the in-circuit CM impedance value in the entire frequency range of 150 kHz to 30 MHz, whereas the VFD's output frequency setting affects the in-circuit CM impedance value in certain frequency regions although this influence is rather small. Future work will study the reasons for the above conclusions through theoretical analysis. In addition, the application of the SPS to extract the in-circuit DM impedance of a MDS will also be explored.

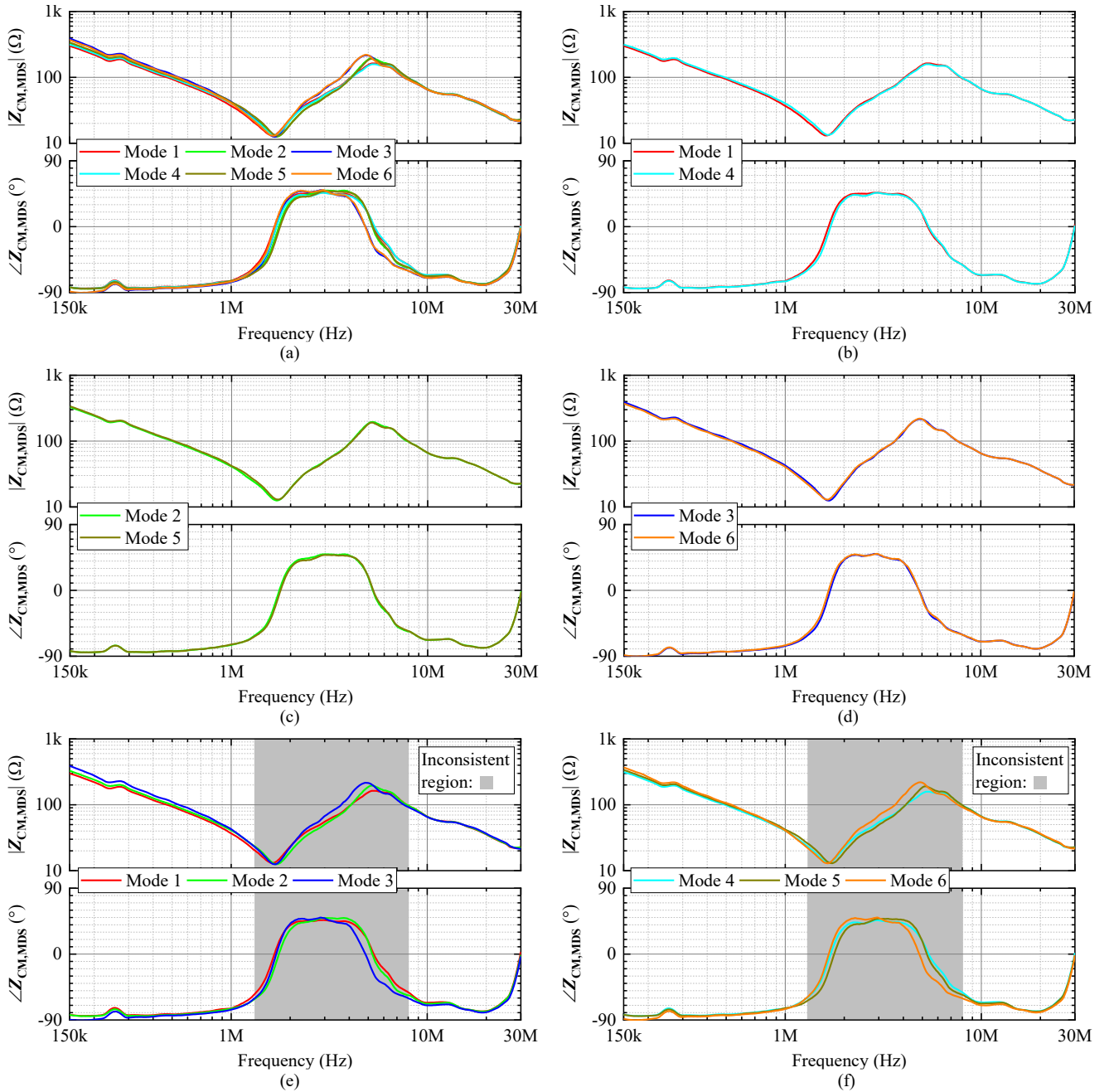

Fig. 6. Measured in-circuit $Z_{CM,MDS}$ from 150 kHz to 30 MHz under: (a) six different operating modes; (b) Mode 1 and Mode 4; (c) Mode 2 and Mode 5; (d) Mode 3 and Mode 6; (e) Mode 1, Mode 2, and Mode 3; (f) Mode 4, Mode 5, and Mode 6.

ACKNOWLEDGMENT

This research work was conducted in the SMRT-NTU Smart Urban Rail Corporate Laboratory with funding support from the National Research Foundation (NRF), SMRT and NTU; under the Corp Lab@University Scheme.

REFERENCES

[1] L. Ran, S. Gokani, J. Clare, K. J. Bradley, and C. Christopoulos, "Conducted electromagnetic emissions in induction motor drive systems. I. Time domain analysis and identification of dominant modes," *IEEE Trans. Power Electron.*, vol. 13, no. 4, pp. 757-767, Jul. 1998, doi: 10.1109/63.704152.

[2] L. Ran, S. Gokani, J. Clare, K. J. Bradley, and C. Christopoulos, "Conducted electromagnetic emissions in induction motor drive systems. II. Frequency domain models," *IEEE Trans. Power Electron.*, vol. 13, no. 4, pp. 768-776, Jul. 1998, doi: 10.1109/63.704154.

[3] V. Tarateeraseth, K. Y. See, F. G. Canavero, and W.-Y. Chang, "Systematic electromagnetic interference filter design based on information from in-circuit impedance measurements," *IEEE Trans. Electromagn. Compat.*, vol. 52, no. 3, pp. 588-597, Aug. 2010, doi: 10.1109/TEMC.2010.2046419.

[4] F. Fan, "Common mode filter design for mitigation of bearing degradation in inverter-fed motor drive systems," Ph.D. thesis, Sch. Electr. Electron. Eng., Nanyang Tech. Univ., Singapore, 2020, doi: 10.32657/10356/138382.

[5] C. J. Kikkert and S. Zhu, "Resistive shunt on-line impedance analyzer," in *Proc. Int. Symp. Power Line Commun. Appl. (ISPLC)*, Bottrop, Germany, 2016, pp. 150-155, doi: 10.1109/ISPLC.2016.7476269.

[6] S. Cobreces, E. J. Bueno, D. Pizarro, F. J. Rodriguez, and F. Huerta, "Grid impedance monitoring system for distributed power generation electronic interfaces," *IEEE Trans. Instrum. Meas.*, vol. 58, no. 9, pp. 3112-3121, Sep. 2009, doi: 10.1109/TIM.2009.2016883.

[7] D. K. Alves, R. Ribeiro, F. B. Costa, and T. Rocha, "Real-time wavelet-based grid impedance estimation method," *IEEE Trans. Ind. Electron.*, vol. 66, no. 10, pp. 8263-8265, Oct. 2019, doi: 10.1109/TIE.2018.2870407.

[8] J. Yao, S. Wang, and H. Zhao, "Measurement techniques of common mode currents, voltages, and impedances in a flyback converter for radiated EMI diagnosis," *IEEE Trans. Electromagn. Compat.*, vol. 61, no. 6, pp. 1997-2005, Dec. 2019, doi: 10.1109/TEMC.2019.2953925.

[9] X. Shang, D. Su, H. Xu, and Z. Peng, "A noise source impedance extraction method for operating SMPS using modified LISN and simplified calibration procedure," *IEEE Trans. Power Electron.*, vol. 32, no. 6, pp. 4132-4139, Jun. 2017, doi: 10.1109/TPEL.2016.2631578.

[10] T. Funaki, N. Phankong, T. Kimoto, and T. Hikihara, "Measuring terminal capacitance and its voltage dependency for high-voltage power devices," *IEEE Trans. Power Electron.*, vol. 24, no. 6, pp. 1486-1493, Jun. 2009, doi: 10.1109/TPEL.2009.2016566.

[11] C. González, J. Pleite, V. Valdivia, and J. Sanz, "An overview of the on line application of frequency response analysis (FRA)," in *Proc. IEEE Int. Conf. Ind. Electron.*, Vigo, Spain, 2007, pp. 1294-1299, 10.1109/ISIE.2007.4374785.

[12] R. A. Southwick and W. C. Dolle, "Line impedance measuring instrumentation utilizing current probe coupling," *IEEE Trans. Electromagn. Compat.*, vol. EMC-13, no. 4, pp. 31-36, Nov. 1971, doi: 10.1109/TEMC.1971.303150.

[13] S. B. Rathnayaka, K. Y. See, and K. Li, "Online impedance monitoring of transformer based on inductive coupling approach," *IEEE Trans. Dielectr. Electr. Insul.*, vol. 24, no. 2, pp. 1273-1279, Apr. 2017, doi: 10.1109/TDEI.2017.006111.

[14] Z. Zhao, K. Y. See, E. K. Chua, A. S. Narayanan, W. Chen, and A. Weerasinghe, "Time-variant in-circuit impedance monitoring based on the inductive coupling method," *IEEE Trans. Instrum. Meas.*, vol. 68, no. 1, pp. 169-176, Jan. 2019.

[15] Z. Zhao, A. Weerasinghe, W. Wang, E. K. Chua, and K. Y. See, "Eliminating the effect of probe-to-probe coupling in inductive coupling method for in-circuit impedance measurement," *IEEE Trans. Instrum. Meas.*, vol. 70, 2021, Art no. 1000908, doi: 10.1109/TIM.2020.3013688.

[16] A. Weerasinghe, Z. Zhao, N. B. Narampanawe, Z. Yang, T. Svimonishvili, and K. Y. See, "Single-probe inductively coupled in-circuit impedance measurement," *IEEE Trans. Electromagn. Compat.*, 2021, doi: 10.1109/TEMC.2021.3091761.

[17] Bode 100 User Manual, OMICRON Lab, Klaus Austria, 2017.